\documentclass[twocolumn,prl,showpacs,floatfix]{revtex4}

\usepackage{subfigure}
\usepackage{graphicx}

\begin{document}

%\draft 

\preprint{DRAFT}

\date{September 17, 2019}

\title{Production and detection of an axion dark matter echo}

\author{Ariel Arza and Pierre Sikivie}

\affiliation{Department of Physics, University of Florida, 
Gainesville, FL 32611, USA}

\begin{abstract}

Electromagnetic radiation with angular frequency equal to half 
the axion mass stimulates the decay of cold dark matter axions 
and produces an echo, i.e. faint electromagnetic radiation 
traveling in the opposite direction.   We propose to search 
for axion dark matter by sending out to space a powerful beam 
of microwave radiation and listening for its echo.  We estimate
the sensitivity of this technique in the isothermal and caustic 
ring models of the Milky Way halo, and find it to be a promising
approach to axion, or axion-like, dark matter detection. 

\end{abstract}
\pacs{95.35.+d}

\maketitle

The identity of dark matter remains one of the central questions 
in science today \cite{dmrev}.  One of the leading candidates is 
the QCD axion.  This hypothetical particle was originally postulated 
as a solution \cite{PQWW} to the Strong CP Problem of the Standard 
Model of particle physics, i.e. the puzzle why the strong interactions 
conserve P and CP.  The properties of the QCD axion are given almost 
entirely in terms of a single parameter $f_a$, called the axion decay 
constant.  In particular the mass of the axion
\begin{equation}
m = 0.6 \times 10^{-5}~{\rm eV}
\left({10^{12}~{\rm GeV} \over f_a}\right)~\ ,
\label{mass}
\end{equation}
and its electromagnetic coupling
\begin{equation}
{\cal L}_{a\gamma\gamma} = - g_\gamma {\alpha \over \pi}
{1 \over f_a} \phi(x) \vec{E}(x) \cdot \vec{B}(x)
\label{agg}
\end{equation}
where $\phi(x)$ is the axion field and $g_\gamma$ is a model-dependent 
dimensionless coupling that is generically of order one.  In the 
KSVZ model \cite{KSVZ}, $g_\gamma = - 0.97$.  In the DFSZ model
\cite{DFSZ}, and in all grand-unified axion models, $g_\gamma = 0.36$.  
Early laboratory limits and stellar evolution constraints require 
$f_a > 10^9$ GeV \cite{Raff3} in which case the axion is so 
extremely weakly coupled that it was once dubbed ``invisible".  
However, cosmology came to the rescue.  Axions are overproduced 
during the QCD phase transition in the simplest cosmological 
scenarios unless $f_a \lesssim 10^{12}$ GeV \cite{axdm}. The 
precise limit depends on whether inflation occurs before or 
after the phase transition during which Peccei-Quinn symmetry 
is spontaneously broken, and other considerations such axion 
production by topological defects, the precise temperature 
dependence of the axion mass, and the amount of entropy 
production associated with the QCD phase transition.  In 
any case, the axions produced during the QCD phase transition 
are a form of cold dark matter \cite{Ipser} and therefore a 
candidate for the constituent particle of galactic 
halos.  The topic is reviewed in refs.~\cite{axrev}.  

Axions and axion-like particles (ALPs) are by-products of 
variously motivated proposals for physics beyond the Standard 
Model, including its supersymmetric extensions \cite{Baer} and 
string theory \cite{Witten}. ALPs \cite{ALPs} are light 
pseudo-scalar particles like QCD axions but without the 
definite relationship between mass and coupling implied 
by Eqs.~(\ref{mass}) and (\ref{agg}).  Several methods 
to test experimentally the axion hypothesis have been 
proposed and some have produced useful limits.  For dark 
matter axion searches, the different approaches include 
the cavity technique \cite{axdet,cavity}, wire \cite{wire} 
and dielectric plate \cite{diel} detectors, magnetic resonance 
methods \cite{NMR,QUAX}, the LC circuit approach \cite{LC}, 
and atomic transitions \cite{atomic}.  Solar axions are 
searched for by their conversion to x-rays in a laboratory 
magnetic field \cite{axdet,solax} and in crystals \cite{Frank}, 
and through the axio-electric effect \cite{axioel}.  "Shining 
light through wall" experiments attempt to convert photons to 
axions on one side of a wall followed by back conversion on 
the other side \cite{SLW}.  Axion induced effects in atoms, 
molecules and nuclei are discussed in ref. \cite{YSVF}.

The purpose of our paper is to propose a new approach to axion 
dark matter detection.   The effect we exploit is the stimulated 
decay of cold dark matter axions by a powerful beam of microwave 
radiation.  Refs. \cite{stimax} discuss the stimulated decay of 
axion dark matter in astrophysical contexts.  We first describe 
the effect in the rest frame of a perfectly cold axion fluid, and 
then generalize to the case where the observer is moving with 
respect to the axion fluid and to the case where the axion 
fluid has velocity dispersion.

Let $\vec{A}_0(\vec{x}, t)$ be the vector potential of the initial 
outgoing radiation.   In the presence of axions, $\vec{A}_0$ is 
itself a source of electromagnetic radiation $\vec{A}_1(\vec{x}, t)$.  
Since axions are very weakly coupled, we have in radiation gauge 
($\vec{\nabla} \cdot \vec{A} = 0$)
\begin{equation}
(\partial_t^2 - \nabla^2) \vec{A}_1 = - g  
(\vec{\nabla} \times \vec{A}_0) \partial_t \phi
\label{weq}
\end{equation}
where $\phi(t) = A \sin(mt)$ is the axion field, and 
$g \equiv g_\gamma {\alpha \over \pi}{1 \over f_a}$ is 
the overall coupling that appears in Eq.~(\ref{agg}).  The 
axion density is $\rho = {1 \over 2} A^2 m^2$.  Let the 
outgoing radiation $\vec{A}_0$ be stationary, linearly 
polarized and with angular frequencies $\omega$ at and 
near $m/2$.  The retarded $\vec{A}_1$ is in that case 
identical to $\vec{A}_0$ except that 1) it flows exactly 
backwards because, up to a constant factor, its spatial 
Fourier transform is the same as that of $\vec{A}_0$ whereas 
its angular frequency is opposite, 2) it is reduced relative 
to $\vec{A}_0$ by a time-dependent factor proportional to 
$g A$, and 3) it is linearly polarized at a $90^\circ$ 
angle relative to $\vec{A}_0$ \cite{Suppl}.  If $\vec{A}_0$ 
is circularly polarized, $\vec{A}_1$ has the same circular 
polarization as $\vec{A}_0$.  We call $\vec{A}_1$ the echo 
wave.

The power in the echo wave is \cite{Suppl}
\begin{equation}
P_1 = {1 \over 16} g^2 \rho {d P_0 \over d \nu} t
\label{echo}
\end{equation}
where ${d P_0 \over d \nu}$ is the spectral density of the 
outgoing wave at angular frequency $\omega = 2 \pi \nu = m/2$, 
and $t$ is the time since the outgoing wave was first established.  
If the outgoing wave is emitted as a parallel beam of finite 
cross-section, it will spread as a result of its transverse 
wavevector components.  The echo wave retraces the outgoing 
wave backward in time, returning to the location of emisssion 
of the outgoing wave with the latter's original transverse size.  
If the outgoing power $P_0$ is turned on for a time $t$ and then 
turned off, the echo power $P_1$ given by Eq.~(\ref{echo}) lasts 
forever in the future under the assumption that the perfectly 
cold axion fluid has infinite spatial extent.  

Next, let us consider the case where the perfectly cold axion 
fluid is moving with velocity $\vec{v}$ with respect to the 
outgoing power source.  Nothing changes in the axion fluid 
rest frame compared to the above discussion except that each 
increment $dE_0 = P_0 dt$ of outgoing energy is emitted from 
a different location.  The incremental echo power $dP_1$, given 
by the RHS of Eq.~(\ref{echo}) with $t$ replaced by $dt$, returns 
forever to the location in the axion fluid rest frame from which 
the increment $dE_0$ of outgoing energy was emitted.  In the 
frame of its source, the frequency at which the outgoing power 
stimulates axion decay is ($c = 1$) 
\begin{equation}
\omega_0 = {m \over 2}(1 + \vec{v} \cdot \hat{k}) + {\cal O}(v^2)
\label{freq0}
\end{equation}
where $\hat{k}$ is the unit vector in the direction of the
outgoing power.  The frequency of the echo is
\begin{equation}
\omega_- = {m  \over 2}(1 - \vec{v} \cdot \hat{k}) + {\cal O}(v^2)~~\ ,
\label{freq-}
\end{equation}
and its direction is $- \hat{k} + 2 \vec{v}_\perp
+ {\cal O}(v^2)$ where $\vec{v}_\perp$ is the component of 
$\vec{v}$ perpendicular to $\hat{k}$.   The echo of power 
emitted a time $t_e$ ago arrives displaced from the point 
of emission by $\vec{d} = \vec{v}_\perp t_e$. To detect as 
much echo power as possible at or near the place of emission 
of the outgoing power, the observer wants $\vec{v}_\perp$ as 
small as possible, i.e. $\hat{k}$ parallel or anti-parallel 
to $\vec{v}$.

If the axion fluid has velocity dispersion, its density can be 
viewed as an integral over cold flows
\begin{equation}
\rho = \int d^3v~{d^3 \rho \over dv^3}(\vec{v})~~\ .
\label{veldist}
\end{equation}
Everything said before holds true for each infinitesimal 
cold flow increment.  The echo frequency has a spread 
$\delta \omega_- = {m \over 2} \delta v_\parallel$ where 
$\delta v_\parallel$ is the spread of axion velocities in 
the $\hat{k}$ direction.  The echo of power emitted a time
$t_e$ ago is spread over a transverse size $\delta \vec{d} 
= \delta \vec{v}_\perp t_e$ where $\delta \vec{v}_\perp$ 
is the spread of axion velocities perpendicular to $\hat{k}$.

It is clear from the above that the amount of echo power
that the observer may easily collect depends sharply on 
the velocity distribution of axion dark matter on Earth 
as well as its total local density.  We will consider
two contrasting models of the Milky Way halo, the
isothermal model \cite{IM} and the caustic ring model
\cite{Duffy}.  In the isothermal model the local dark 
matter has density 300 MeV/cm$^3$ and velocity dispersion 
270 km/s. In the caustic ring model, the local dark 
matter velocity distribution is dominated by a single 
flow, called the Big Flow, because of our proximity to 
the 5th caustic ring of dark matter in the Milky Way halo.
An upper limit of order 70 m/s on the velocity dispersion 
of the Big Flow has been derived \cite{veldis}. The 
direction of the Big Flow can be obtained from triangular
features in the IRAS and GAIA maps of the Galactic plane
\cite{CGHS} with a precision of order 0.01 rad.  The 
density of the Big Flow is poorly constrained because 
it depends sharply on our distance and position relative 
to the mid-plane cusp of the nearby caustic ring.  It is 
1 GeV/cm$^3$ \cite{Duffy} at least, but may be as large 
as 10 GeV/cm$^3$ or even higher \cite{CGHS}.  Such high 
densities pertain only to a small region within 10 pc or 
so of the aforementioned cusp and are consistent with 
measurements of the Galactic rotation curve and of the 
vertical motions of stars in the solar neighborhood.

Let us first consider the case where the local axion 
dark matter density is dominated by a single cold
flow, as in the caustic ring model.  As was discussed 
above, the largest amount of echo power is available 
for detection near the source of outgoing power when 
the outgoing power has direction $\hat{k}$ parallel 
or anti-parallel to the velocity vector $\vec{v}$ of 
the axion fluid with respect to the observer.  That 
velocity vector is a sum
\begin{equation}
\vec{v}(t) = \vec{v}_a - \vec{v}_{\rm LSR}
-\vec{v}_\odot - \vec{v}_\otimes(t) 
\label{BFv}
\end{equation}
where $\vec{v}_a$ is the velocity of the axion fluid with 
respect to a non-rotating coordinate system attached to the 
Milky Way galaxy, $\vec{v}_{\rm LSR}$ is the velocity of the 
Local Standard of Rest (LSR) in that same coordinate system, 
$\vec{v}_\odot$ is the velocity of the Sun with respect to the 
LSR, and $\vec{v}_\otimes$ is the velocity of the observer with 
respect to the Sun as a result of the orbital and rotational 
motions of the Earth.  We are particularly interested in the 
extent to which the uncertainties in the several terms on the RHS 
of Eq.~(\ref{BFv}) affect our ability to minimize $\vec{v}_\perp$.
$\vec{v}_\otimes(t)$ is known with great precision.  The components 
of $\vec{v}_\odot$ are known with a precision of order 3 km/s.  
$\vec{v}_{\rm LSR}$ is in the direction of Galactic rotation by 
definition.  Its magnitude (often quoted to be 220 km/s) is known 
with an uncertainty of order 20 km/s.  The magnitude of $\vec{v}_a$ 
for the Big Flow is approximately 520 km/s \cite{Duffy}.  Its 
direction is fixed by the positions of the IRAS and GAIA 
triangles on the sky with a precision of order 0.01 radians
\cite{CGHS}.  So we expect that it is not possible to reduce 
$\vec{v}_\perp$ to less than of order 5 km/s, the nominal value 
we use below.  Because the Big Flow is almost in the direction 
of Galactic rotation (within approximately $12^\circ$), the 
uncertainty on the magnitude of $\vec{v}_{\rm LSR}$ is less 
important.

Consider a dish (e.g. a radiotelescope) of radius $R$
collecting echo power at or near the location of the 
outgoing power source.  Because the echo from outgoing 
power emitted a time $t_e$ ago is displaced by 
$\vec{d} = \vec{v}_\perp t_e$, the amount of echo 
power collected by the dish is  
\begin{equation}
P_c = {1 \over 16} g^2 \rho {d P_0 \over d \nu} 
C {R \over |\vec{v}_\perp|}
\label{echo2}
\end{equation}
where $C$ is a number of order one which depends
on the configuration of the source relative to 
the receiver dish:
\begin{equation}
C = {|\vec{v}_\perp| \over 2 R P_0} \int dt 
\int_{S_0}  d^2x ~I_0(\vec{x}) 
\Theta_c(\vec{x} + \vec{v}_\perp t)~~\ .
\label{C}
\end{equation}
Here $S_0$ is the surface from which the outgoing 
power is emitted, $I_0(\vec{x})$ is the outgoing 
power per unit surface, and $\Theta_c(\vec{x})$ is 
a function that equals one if $\vec{x}$ belongs
to the receiver dish area and zero otherwise.
For example, $C = 0.5$ if the outgoing power 
is emitted from the center of the receiver dish, 
whereas $C = 0.424$ if the outgoing power is 
emitted uniformly from the area of the receiver 
dish.  However neither of these configurations 
is likely to be optimal.  It is probably better 
to place several source dishes around the receiver 
dish. $C$ can be straightforwardly calculated for 
each configuration.

Let us assume that a pulse of outgoing power $P_0$, 
with frequency $\nu_0$ and uniform 
spectral density ${dP_0 \over d\nu} = {P_0 \over \Delta \nu_0}$
over bandwidth $\Delta \nu_0$, is emitted during a time $t_m$.  
Provided that 
\begin{equation}
t_m > {R \over 2 |\vec{v}_\perp|} 
= 0.5 \times 10^{-2}~{\rm sec}~{R \over 50~{\rm m}}~
{5~{\rm km/s} \over |\vec{v}_\perp|}~~\ ,
\label{cond}
\end{equation}
the echo power  
\begin{eqnarray}
P_c &=& 2.33 \times 10^{-31} P_0 
\left({10~{\rm kHz} \over \Delta \nu_0}\right)
\left({g_\gamma \over 0.36}\right)^2
\left({10^{12}~{\rm GeV} \over f_a}\right)^2 \cdot\nonumber\\
&\cdot& \left({\rho \over {\rm GeV/cm}^3}\right)
\left({C \over 0.30}\right)
\left({R \over 50~{\rm m}}\right)
\left({5~{\rm km/s} \over |\vec{v}_\perp|}\right)~~\ ,
\label{PC}
\end{eqnarray}
is received over the same time interval $t_m$. Since the 
magnitude of the velocity of the Big Flow relative to us 
$v \simeq$ 520 km/s - 220 km/s = 300 km/s, the frequency 
of the echo power is red- or blue-shifted from $\nu_0$ by 
$\Delta \nu \simeq 2 \times 10^{-3} \nu_0$.  The echo 
power has bandwidth $B = 2 \delta v~\nu < 5 \times 10^{-7} \nu$ 
since the velocity dispersion of the Big Flow is less than 
70 m/s \cite{veldis}.  The frequency range of interest is 
approximately 30 MHz to 30 GHz because the Earth's 
atmosphere is mostly transparent at those frequencies. 
It corresponds to the mass range $2.5 \times 10^{-7} < 
m < 2.5 \times 10^{-4}$ eV, which happens to be prime 
hunting ground for QCD axions.
  
The cosmic microwave background and radio emission by 
astrophysical sources are irreducible sources of noise.
In addition there is instrumental noise.  The total noise 
temperature depends on frequency, on the location of the 
telescope and on the direction of observation.  As an example 
we may consider the system noise temperature of the Green 
Bank Telescope \cite{GBT}: approximately 20 K from 1 GHz 
to 8 GHz, approximately linearly rising from 20 K at 
8 GHz to 40 K at 30 GHz, and exponentially rising 
towards low frequencies from 20 K at 1 GHz to 100 K 
at 300 MHz.  The rise at low frequencies is due
to Galactic emission and is strongly direction 
dependent.  100 K at 300 MHz is a typical value. 
The rise at high frequencies is due to atomic and 
molecular transitions in the atmosphere.  It can 
be mitigated by placing the telescope at a high 
elevation.  We use below a nominal system noise 
temperature of 20 K at all frequencies for the 
purpose of stating the results of our sensitivity 
calculations. 

The signal to noise ratio with which the echo power 
is detected when $\omega_0$ falls within the angular 
frequency range of the emitted power is given by Dicke's 
radiometer equation
\begin{equation}
s/n = {P_c \over T_n} \sqrt{t_m \over B}~~\ .
\label{Dicke}
\end{equation}
Combining Eqs.~(\ref{PC}) and (\ref{Dicke}) and 
setting $B = 5 \times 10^{-7} \nu$, the total outgoing 
energy per logarithmic frequency interval necessary to 
detect the axion echo with a given signal to noise ratio
is found to be:  
\begin{eqnarray}
{d E_0 \over d \ln \nu}\Bigg|_{\rm BF} &=& 7.2~{\rm MW year} 
\left({s/n \over 5}\right) 
\left({10~{\rm GHz} \over \nu}\right)^{1 \over 2}
\left({0.36 \over g_\gamma}\right)^2
\cdot\nonumber\\
&\cdot& 
\left({T_n \over 20~{\rm K}}\right)
\left({{\rm GeV/cm}^3 \over \rho}\right)
\left({0.30 \over C}\right)
\cdot\nonumber\\
&\cdot&
\left({t_m \over 10^{-2}~{\rm sec}}\right)^{1 \over 2}
\left({50~{\rm m} \over R}\right)
\left({|v_\perp| \over 5~{\rm km/s}}\right)\ .
\label{outpow}
\end{eqnarray}
We used Eq.~(\ref{mass}) and $m = 4 \pi \nu$.  Fig. 1 shows 
the sensitivity to $g_\gamma$ of an axion echo search that 
consumes 10 MWyear of outgoing energy for each octave (factor 
of 2) in axion mass covered, for $\rho$ = 1 and 10 GeV/cm$^3$ 
and the nominal values of all other experimental parameters 
used in Eq.~(\ref{outpow}).  

In the isothermal model, $\vec{v}_a$ = 0 on average.  In a 
non-rotating Galactic reference frame the velocity distribution 
is Gaussian with dispersion $\sqrt{3} \sigma \equiv 
\sqrt{\langle \vec{v}\cdot\vec{v} \rangle} \simeq$ 270 km/s.  
In the LSR, the axion fluid moves with speed 220 km/s in the 
direction opposite to that of Galactic rotation.  Assuming 
the direction $\hat{k}$ of the outgoing power is parallel 
(anti-parallel) to the direction of Galactic rotation the 
echo power is blue (red)-shifted in frequency by a fractional 
amount whose average is
$\langle {\Delta \nu \over \nu} \rangle \simeq$ 440 km/s 
= $1.5 \times 10^{-3}$ and whose rms deviation is
${\delta \nu \over \nu} = 2 \sigma  \simeq 1.04 \times 10^{-3}$.  
The echo from outgoing energy that was emitted a time $t_e$ 
ago is spread in space over a fuzzy circular region whose 
radius is Gaussian-distributed with average value $\sigma t_e$.  
Eq.~(\ref{echo2}) holds with 
${1 \over |\vec{v}_\perp|}$ replaced by 
\begin{equation}
\langle {1 \over |\vec{v}_\perp|} \rangle
= \sqrt{\pi \over 2} {1 \over \sigma} = 
{1 \over 124~{\rm km/s}}~~\ .
\label{sub}
\end{equation}
In view of Eq.~(\ref{cond}) we now require 
$t_m > 2 \times 10^{-4} {\rm sec} {R \over 50~{\rm m}}$.
Using Eq.~(\ref{Dicke}) with $B = 4 \sigma \nu
= 2.1 \times 10^{-3} \nu$ and setting $\rho$ = 0.3 GeV/cm$^3$, 
we find 
\begin{eqnarray}
{d E_0 \over d \ln \nu}\Bigg|_{\rm iso} = 5.3~{\rm GWyear}
\left({s/n \over 5}\right)
\left({10~{\rm GHz} \over \nu}\right)^{1 \over 2}
\left({0.36 \over g_\gamma}\right)^2
\cdot\nonumber\\
\cdot
\left({T_n \over 20~{\rm K}}\right)
\left({0.30 \over C}\right)
\left({t_m \over 2 \cdot 10^{-4}~{\rm sec}}\right)^{1 \over 2}
\left({50~{\rm m} \over R}\right)\ .
\label{outpowi}
\end{eqnarray}
The sensitivity to $g_\gamma$ in the isothermal model 
is shown in Fig. 1 as well.

\begin{figure}
\includegraphics[width=0.9\columnwidth]{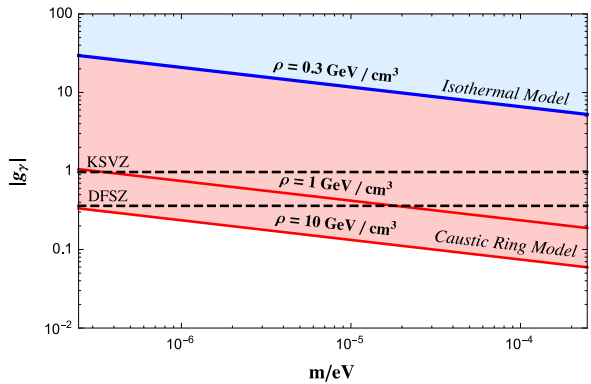}
\caption{Sensitivity to $|g_\gamma|$ as function of 
mass of an axion echo search consuming 10 MWyear 
of outgoing energy for each factor two in axion mass
range covered, in the caustic ring model with Big Flow
densities $\rho$ = 1 and 10 GeV/cm$^3$, and in the 
isothermal model with density $\rho$ = 0.30 GeV/cm$^3$, 
assuming all other experimental parameters have the 
nominal values shown in Eqs.~(\ref{outpow}) and 
(\ref{outpowi}).}
\label{plot1}
\end{figure}

The echo method appears an attractive approach to axion 
dark matter detection because it uses relatively old 
technology and because it is applicable over a wide 
range of axion masses.  The method works better in the 
caustic ring model than in the isothermal model for 
three reasons: 1) the density is higher, 2) the echo 
has less spread in frequency, and 3) the echo has less 
spread in physical space.  Higher density helps the 
cavity method equally.  Fig. 2 shows the sensitivity 
of the echo method to $|g_\gamma| \sqrt{\rho}$ as well 
as the published limits obtained by searches using the
cavity method.  The small spread in frequency of a 
signal from a cold flow, such as the Big Flow of the 
caustic ring model, also helps somewhat the cavity 
method.  Ref. \cite{Hires} describes a high resolution 
analysis of ADMX data and shows that an improvement 
by a factor of order 2 in $g_\gamma \sqrt{\rho}$ is 
obtained in case of a cold flow of velocity dispersion
less than 10 m/s.  The improvements in sensitivity of 
the cavity method in case of cold flows are not included 
in the published bounds shown in Fig. 2.

\begin{figure}
\includegraphics[width=0.9\columnwidth]{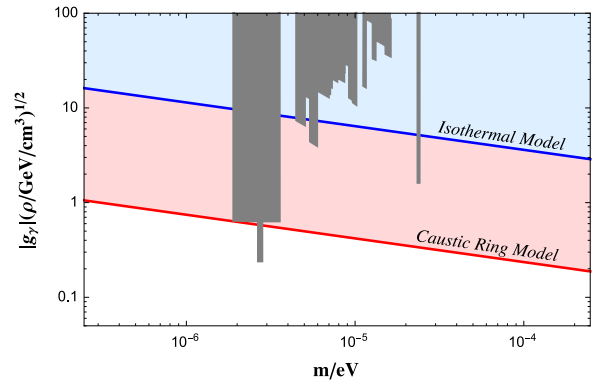}
\caption{Sensitivity to $|g_\gamma| \sqrt{\rho}$ as 
a function of mass of an axion echo search consuming
10 MWyear of outgoing energy for each factor two in 
axion mass range searched, in the caustic ring model 
and in the isothermal model, assuming the nominal 
values of the experimental parameters shown in 
Eqs.~(\ref{outpow}) and (\ref{outpowi}).  The grey 
shaded areas are ruled out by searches using the cavity 
method.}
\label{plot2}
\end{figure}

We thank Richard Bradley, Fritz Caspers, Guido Mueller, 
Neil Sullivan, David Tanner, Qiaoli Yang and Konstantin 
Zioutas for useful discussions.  This research was 
supported in part by the U.S. Department of Energy 
under grant DE-SC0010296, by the Chilean Commission 
on Research, Science and Technology (CONICYT) under 
grant 78180100 (Becas Chile, Postdoctorado), and by 
the Heising-Simons Foundation under grant No. 2015-109.


\begin{thebibliography}{}

\bibitem{dmrev}
Reviews include: {\it Particle Dark Matter} edited by
Gianfranco Bertone, Cambridge University Press 2010;
E.W. Kolb and M. Turner, {\it The Early Universe},
Addison Wesley 1990.

\bibitem{PQWW}
R. D. Peccei and H. Quinn, Phys. Rev. Lett. {\bf 38} (1977) 1440 
and Phys.Rev. {\bf D16} (1977) 1791; S. Weinberg, Phys. Rev. Lett. 
{\bf 40} (1978) 223; F. Wilczek, Phys. Rev. Lett. {\bf 40} (1978) 
279.

\bibitem{KSVZ}
J. Kim, Phys. Rev. Lett. {\bf 43} (1979) 103; M. A. Shifman,
A. I. Vainshtein and V. I. Zakharov, Nucl. Phys. {\bf B166}
(1980) 493.

\bibitem{DFSZ}
M. Dine, W. Fischler and M. Srednicki,
Phys. Lett. {\bf B104} (1981) 199;
A. Zhitnitskii, Sov. J. Nucl. 31 (1980) 260.

\bibitem{Raff3}
G. Raffelt, Lect. Notes Phys. D79 (2008) 51.

\bibitem{axdm}
J. Preskill, M. Wise and F. Wilczek, Phys. Lett. B120 (1983) 127;
L. Abbott and P. Sikivie, Phys. Lett. B120 (1983) 133;
M. Dine and W. Fischler, Phys. Lett. B120 (1983) 137.

\bibitem{Ipser}
J. Ipser and P. Sikivie, Phys. Rev. Lett. 50 (1983) 925.

\bibitem{axrev}
P. Sikivie, Lect. Notes Phys. 741 (2008) 19.
D.J.E. Marsh, Phys. Rep. 643 (2016) 1.

\bibitem{Baer}
H. Baer, V. Barger and D. Sengupta, Phys. Lett. B790 (2019) 58, 
and references therein.

\bibitem{Witten}
P. Svrcek and E. Witten, JHEP 0606 (2006) 051.

\bibitem{ALPs}
P. Arias et al., JCAP 1206 (2012) 013, and references therein.

\bibitem{axdet}
P. Sikivie, Phys. Rev. Lett. 51 (1983) 1415 and
Phys. Rev. D32 (1985) 2988.

\bibitem{cavity}
S.J. Asztalos et al., Phys. Rev. Lett. 104 (2010) 041301;
N. Du et al., Phys. Rev. Lett. 120 (2018) 151301;
B.M. Brubaker et al., Phys. Rev. Lett. 118 (2017) 061302;
S. Youn, Int. J. Mod. Phys. Conf. Ser. 43 (2016) 1660193;
B.T. McAllister et al., Phys. Dark Univ. 18 (2017) 67.

\bibitem{wire}
P. Sikivie, D.B. Tanner and Y. Wang, Phys. Rev. D50 (1994) 4744;
G. Rybka and A. Wagner, Phys. Rev. D91 (2015) 011701.

\bibitem{diel}
A. Caldwell et al., Phys. Rev. Lett. 118 (2017) 091801.

\bibitem{NMR}
P. Graham and S. Rajendran, Phys. Rev. D88 (2013) 035023;
D. Budker et al., Phys. Rev. X 4 (2014) 021030;
P. Graham et al., Phys. Rev. D97 (2018) 055006.

\bibitem{QUAX}
R. Barbieri et al., Phys. Dark Univ. 15 (2017) 135.

\bibitem{LC}
P. Sikivie, N. Sullivan and D.B. Tanner, Phys. Rev. Lett. 112
(2014) 131301; Y. Kahn, B. Safdi and J. Thaler, Phys. Rev. Lett.
117 (2016) 141801; B.T. McAllister, S.R. Parker and M.E. Tobar, 
Phys. Rev. D94 (2016) 042001;  N. Crisosto et al., Springer Proc. 
Phys. 211 (2018) 127; J.L. Ouellet et al., Phys. Rev. Lett. 
122 (2019) 121802.

\bibitem{atomic}
P. Sikivie, Phys. Rev. Lett. 113 (2014) 201301;
C. Braggio et al., Sci. Rep. 7 (2017) 15168;
V.V. Flambaum et al., J. Mod. Phys. A33 (2018) 1844030.

\bibitem{solax}
Y. Inoue et al., Phys. Lett. B668 (2008) 93;
V. Anastassopoulos et al., Nature Physics 13 (2017) 584;  
E. Armengaud et al., JINST 9 (2014) T05002.

\bibitem{Frank}
F. Avignone et al., Phys. Rev. Lett. 81 (1998) 5068.

\bibitem{axioel}
S. Dimopoulos, G.D. Starkman and B.W. Lynn,
Phys. Lett. B168 (1986) 145;  D.S. Akerib et al., 
Phys. Rev. Lett. 118 (2017) 261301, and references therein.

\bibitem{SLW}
K. van Bibber et al., Phys. Rev. Lett. 59 (1987) 759;
K. Ehret et al., Phys. Lett. B689 (2010) 149, and references
therein; R. Ballou et al., Phys. Rev. D92 (2015) 092002;
F. Hoogeveen and T. Ziegenhagen, Nucl. Phys. B358 (1991) 3;
G. Mueller, P. Sikivie, D. Tanner and K. van Bibber,
Phys. Rev. Lett. D80 (2009) 072004.

\bibitem{YSVF}
Y.V. Stadnik and V.V. Flambaum, Phys. Rev. D89 (2014) 043522.

\bibitem{stimax}
I.I. Tkachev, JETP Lett. 101 (2015) 1; 
A. Arza, Eur. Phys. J. C79 (2019) 250; 
A. Caputo et al., JCAP 1903 (2019) 27;
A. Caputo, C. Pe\~na Garay and S.J. Witte, 
Phys. Rev. D98 (2018) 083024 and Erratum:
Phys. Rev. D99 (2019) 089901.

\bibitem{Suppl}
See Supplemental Material at [URL ...] for a derivation
of these results.

\bibitem{IM}
M.S. Turner, Phys. Rev. D33 (1986) 889.

\bibitem{Duffy}
L. Duffy and P. Sikivie, Phys. Rev. D78 (2008) 063508.

\bibitem{veldis}
N. Banik and P. Sikivie, Phys. Rev. D93 (2016) 103509.

\bibitem{CGHS}
S. Chakrabarty, A. Gonzales, Y. Han and P. Sikivie, to be 
published.

\bibitem{GBT}
https://science.nrao.edu/facilities/gbt/proposing/GBTpg.pdf .

\bibitem{Hires}
L. Duffy et al., Phys. Rev. Lett. 95 (2005) 091304 and 
Phys. Rev. D74 (2006) 012006.

\end{thebibliography}
\end{document}